\documentclass[letterpaper]{article} % DO NOT CHANGE THIS
\usepackage{aaai2026}  % DO NOT CHANGE THIS
\usepackage{times}  % DO NOT CHANGE THIS
\usepackage{helvet}  % DO NOT CHANGE THIS
\usepackage{courier}  % DO NOT CHANGE THIS
\usepackage[hyphens]{url}  % DO NOT CHANGE THIS
\usepackage{graphicx} % DO NOT CHANGE THIS
\usepackage{natbib}  % DO NOT CHANGE THIS AND DO NOT ADD ANY OPTIONS TO IT
\usepackage{caption} % DO NOT CHANGE THIS AND DO NOT ADD ANY OPTIONS TO IT
\usepackage{algorithm}
\usepackage{algorithmic}

\usepackage{newfloat}
\usepackage{listings}
\DeclareCaptionStyle{ruled}{labelfont=normalfont,labelsep=colon,strut=off} % DO NOT CHANGE THIS
\floatstyle{ruled}
\newfloat{listing}{tb}{lst}{}
\floatname{listing}{Listing}
\nocopyright 
\title{PyBatchRender: A Python Library for Batched 3D Rendering at Up to One Million FPS}

\author{
    Evgenii Rudakov\textsuperscript{\rm 1},
    Jonathan Shock\textsuperscript{\rm 2, 3, 4, 5},
    Benjamin Ultan Cowley\textsuperscript{\rm 1, 6}
}
\affiliations{
  \textsuperscript{\rm 1}Faculty of Educational Sciences, University of Helsinki\\
  \textsuperscript{\rm 2}Department of Mathematics and Applied Mathematics, University of Cape Town\\
  \textsuperscript{\rm 3}Neuroscience Institute, University of Cape Town\\
  \textsuperscript{\rm 4}INRS, Montreal, Canada\\
  \textsuperscript{\rm 5}NiTheCS, Stellenbosch, South Africa\\
  \textsuperscript{\rm 6}Cognitive Science, Department of Digital Humanities, University of Helsinki\\
    evgenii.rudakov@helsinki.fi, jonathan.shock@uct.ac.za, ben.cowley@helsinki.fi
}

\usepackage{bibentry}
\begin{document}

\maketitle

\begin{abstract}
Reinforcement learning from pixels is often bottlenecked by the performance and complexity of 3D rendered environments.
Researchers face a trade-off between high-speed, low-level engines and slower, more accessible Python frameworks. To address this, we introduce PyBatchRender, a Python library for high-throughput, batched 3D rendering that achieves over 1 million FPS on simple scenes. Built on the Panda3D game engine, it utilizes its mature ecosystem while enhancing performance through optimized batched rendering for up to 1000× speedups. Designed as a physics-agnostic renderer for reinforcement learning from pixels, PyBatchRender offers greater flexibility than dedicated libraries, simpler setup than typical game-engine wrappers, and speeds rivaling state-of-the-art C++ engines like Madrona. Users can create custom scenes entirely in Python with tens of lines of code, enabling rapid prototyping for scalable AI training. Open-source and easy to integrate, it serves to democratize high-performance 3D simulation for researchers and developers. The library is available at 
\url{https://github.com/dolphin-in-a-coma/PyBatchRender}.

% It's optimized for use on Nvidia GPUs as well as on Mac ARM devices.
% is ecosystem the right word here, or rather functional?
\end{abstract}

\begin{figure*}[t]
  \centering
  \includegraphics[width=\textwidth]{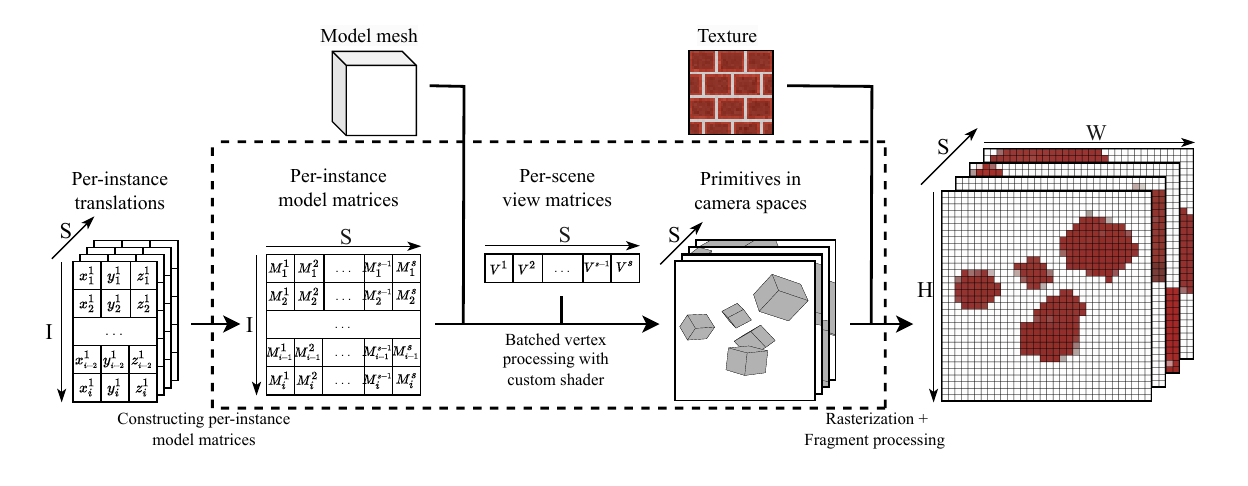} % jpg/png/pdf only
  \caption{Batched rendering pipeline for a single geometry across multiple scenes. Users specify per-instance translations for $I$ instances of a shared mesh across $S$ scenes (dashed box separates user control from internal operations). The system constructs per-instance model matrices $M_i^s$, composes them with per-scene view matrices $V^s$, and processes all instances across all scenes in a single pass using a custom batched vertex shader. Rasterization and fragment processing produce a tiled output texture, which is then partitioned into $S$ individual frames of resolution $H \times W$ that remain on the GPU for subsequent access. This approach consolidates $S \times I$ independent render passes into a single parallelized GPU operation, eliminating render target switching overhead and maximizing hardware utilization. \emph{Not shown:} per-instance \emph{scales}, \emph{rotations}, and \emph{colors}, as well as camera control via e.g. positions/orientations, all of which are supported by the system but omitted here for clarity.}
  \label{fig:teaser}
\end{figure*}

\section{Introduction}

Reinforcement learning (RL) from pixels represents a critical area with substantial real-world applicability, including research on embodied AI and video games. However, training directly from visual observations presents significant challenges compared to vector-based environments. Pixel-based methods exhibit substantially lower sample efficiency, requiring orders of magnitude more training data than direct vector observations \cite{yarats_improving_2019}. The rendering costs compound this challenge. Each pixel observation demands significantly more computational resources than state-based observations. This performance gap is particularly severe for complex 3D rendering scenarios, where rendering to pixels can be 100 times slower than physics steps \cite{zakka_mujoco_2025}. These challenges call for optimized rendering.

Until recently, the problem of specialized data-oriented rendering has not been solved, making it the primary bottleneck in RL from pixels. Various efforts have been made to optimize rendering for data-intensive pipelines, but they have not achieved widespread adoption due to their complexity \cite{shacklett_large_2021, berges_galactic_2023}. 

Some simulators have extended physics engines with rendering capabilities, achieving modest visual throughput. Isaac Lab (built on NVIDIA's Isaac Sim) \cite{mittal_orbit_2023} and ManiSkill3 (built on the open-source SAPIEN framework) \cite{mu_maniskill_2021, tao_maniskill3_2025} represent GPU-parallelized approaches that integrate physics simulation with rendering. Both frameworks can reach speeds up to 50k FPS in simple environments. However, they remain tightly coupled to their physics engines and optimized primarily for robotics manipulation rather than general-purpose rendering. 

Madrona represents a significant advance in high-throughput rendering \cite{shacklett_extensible_2023,rosenzweig_high-throughput_2024}. Recently, Madrona rendering was integrated in MuJoCo Playground \cite{zakka_mujoco_2025}, achieving 403k steps per second on benchmarks like 3D CartPole. While Madrona delivers impressive performance, it is relatively complex. Madrona requires environment specifications in C++, posing a barrier for RL in practice, where workflows tend to be Python-based. Notably, Madrona and its predecessors also selected the approach of developing an RL-oriented environment from the ground-up, without relying on the convenience of established engines.

Traditional game engines have also been adapted for RL. Platforms like Unity (with ML-Agents) \cite{noauthor_unity-technologiesml-agents_2025} and Godot (with godot-rl) \cite{beeching_godot_2021} offer rich rendering and physics. While more accessible than specialized renderers, these solutions require a user to establish communication protocols between the game engine and training scripts. This creates a communication bottleneck, often limiting simulation throughput to 2k-4k frames per second (FPS) for simple environments \cite{beeching_godot_2021}.

Panda3D, a Python-oriented game engine \cite{goslin_panda3d_2004}, has gained traction for its accessibility among Python developers. Its flexibility has made it the foundation for complex driving simulators like MetaDrive \cite{li_metadrive_2023} and a number of various scientific simulators \cite{kiefl_pooltool_2024, goslin_panda3d_2004}. Benefits include low-level control via shaders and a lack of rigid structure. Nevertheless, Panda3D was not originally designed for massively batched rendering, resulting in modest throughput (often 1k-2k FPS \cite{li_metadrive_2023}) in typical RL training loops.

This landscape presents a clear trade-off: researchers must choose between the high performance but high complexity of specialized C++ engines, and the accessibility but low performance of traditional game engine wrappers.

To bridge the gap between ease of use and high performance, we present PyBatchRender, a data-oriented 3D rendering library. Our library is designed for simple installation, modular environment specification in pure Python, and high-throughput performance. It delivers over 1 million frames per second on simple-geometry environments on both data-center and consumer GPUs.

\section{Methods}
PyBatchRender's approach is different from previous data-generation and RL-oriented renderers. Rather than developing a rendering engine from scratch, our library is built as a performance-focused interface over the mature Panda3D game engine. This architectural decision allows us to focus on the critical bottlenecks in batched rendering, including batched scene control and pixel data acquisition. At the same time we delegate traditional rendering operations to Panda3D's already-optimized implementations. As an additional benefit, this approach enables straightforward adaptation of existing Panda3D features, including asset loading and shader pipelines, into the high-throughput rendering context.

Panda3D provides a comprehensive foundation for 3D rendering, handling model loading, texture management, rasterization, and scene graph traversal. Beyond these high-level functions, it exposes low-level control through custom GLSL shaders, allowing fine-grained manipulation of the way in which individual instances are drawn. However, Panda3D's API was designed for traditional game development, where objects and cameras are manipulated individually through Python function calls. This design incurs substantial overhead when managing thousands of parallel environments, as each object requires separate API calls that cannot be efficiently vectorized.

We address this limitation through a set of targeted extensions that transform Panda3D into a flexible high-throughput, batched rendering system suitable for reinforcement learning from pixels. These extensions include: (1) multi-view tiled rendering, (2) tensor-based batched control of instanced geometry through custom shaders, (3) CUDA-OpenGL interoperability, and (4) integration with TorchRL.  We now detail each of these components and their role in sequence.

\subsubsection{Multi-View Tiling}
This technique renders $K$ independent scenes (e.g., $K$ parallel RL environments) into a single, large offscreen render target. The target is partitioned into a grid of non-overlapping tiles, and each scene is assigned its own view-projection matrix. A lightweight viewport remapping operation in the vertex shader projects each scene's geometry into its designated tile. This reduces engine and driver overhead by consolidating $K$ render passes into one.

\subsubsection{Hardware-Instanced Geometry Control}
We bypass the Python API for object control by using hardware instancing. PyBatchRender exposes an interface where users control entire batches of independent objects using single tensor commands, common in PyTorch and NumPy. For $S$ scenes and $I$ instances of a given model, the user provides tensors for transforms (e.g., $S \times I \times 3$ positions, $S \times I \times 3$ orientations, $S \times I \times 1$ scales) and appearance (e.g., $S \times I \times 4$ colors).

Our library vectorially composes these user-submitted tensors into $4 \times 4$ model matrices, which are packed into buffer textures and consumed by a custom GLSL shader. This architecture issues a single instanced draw call per unique model type, allowing the GPU to render all instances across all scenes in parallel without any Python-loop overhead. This technique also allows instance transforms to be shared across scenes, reducing bandwidth in scenarios where many objects are shared between camera views, e.g. in multi-agent reinforcement learning. 

\subsubsection{CUDA-OpenGL Interoperability}
For NVIDIA GPUs, we implement a direct CUDA–OpenGL interoperability path. This process maps the final OpenGL output texture directly into CUDA memory space. A device-to-device copy then transfers the pixels into a pre-allocated GPU buffer, which is exposed to PyTorch via the DLPack protocol. This entire operation avoids any `host round-trip' (GPU $\rightarrow$ CPU $\rightarrow$ GPU), eliminating the PCIe bus as a data-transfer bottleneck.

To ensure broad accessibility, we also provide a portable CPU path. This fallback method works on any hardware, including AMD GPUs and Apple Silicon. It performs a standard texture extraction to a contiguous buffer in system RAM, which is then returned as a PyTorch tensor. While subject to PCIe transfer limits, this path ensures full functionality for development and execution on non-NVIDIA systems.

\subsubsection{System Integration}
PyBatchRender is intentionally physics-agnostic. It is designed as a modular rendering component. The library exposes a simple interface where physics engines update object states (positions, orientations, scales) as tensors, and PyBatchRender handles the visualization. This separation of concerns allows researchers to swap physics backends or rendering configurations independently.

To facilitate integration into RL training pipelines, PyBatchRender provides a TorchRL wrapper that implements the standard Gymnasium environment interface. This integration enables seamless use of TorchRL's parallel environment utilities, which support multi-process vectorization without requiring data to be copied from the GPU to the CPU for inter-process communication.

\section{Results}

% We perform a series of ablation studies, shown in Table 3, where we cumulatively add components until we reach the full model. 

We conduct two primary experiments to evaluate the performance of PyBatchRender. First, a cumulative performance gain analysis validates the performance contribution of each of our core architectural optimizations. Second, we benchmark PyBatchRender against other state-of-the-art, high-throughput rendering libraries.

To validate the effectiveness of our design choices, we progressively add each optimization to a baseline renderer. All measurements use the CartPoleBalance environment rendered at 64×64 resolution across three representative GPU architectures: Apple M2 (representing modern unified-memory laptops), NVIDIA RTX 4090 (high-end consumer GPU), and NVIDIA A100 (data-center class GPU). These platforms span distinct performance profiles in terms of memory bandwidth, rasterization capabilities, and CPU-GPU communication overhead.

\begin{figure}
    \centering
    \includegraphics[width=1\linewidth]{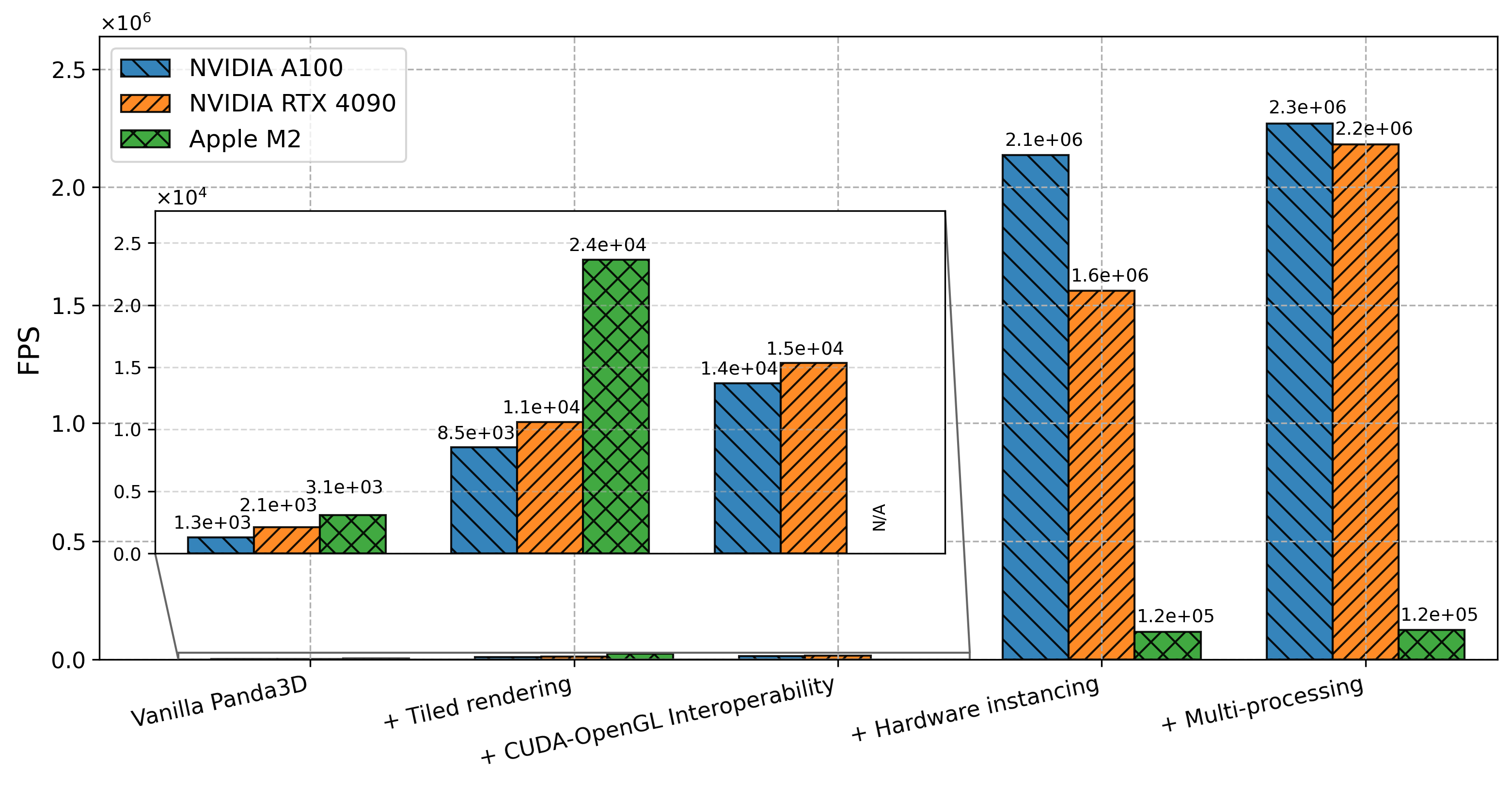}
    \caption{Cumulative performance gains for rendering throughput for 64x64 resolution in the CartPoleBalance environment. Each component is added cumulatively to the baseline, demonstrating its contribution to the final performance on Apple M2, Nvidia RTX 4090, and Nvidia A100 GPUs.}
    \label{fig:ablation}
\end{figure}

Figure~\ref{fig:ablation} presents the cumulative performance gains as components are added to a naive Panda3D baseline implementation. Our baseline, a naive Panda3D implementation, achieves 1.1k–3.3kFPS. The M2 shows the strongest baseline performance, likely benefiting from its unified memory architecture, which avoids PCIe bus latency. The A100, optimized for general compute over rasterization, shows the lowest baseline result.

Introducing tiled rendering, through independent scenes managed iteratively in Python, yields immediate benefits. The RTX 4090, A100, and M2 improve by factors of 5.2×, 6.5×, and 7.8× respectively. 
This stage consolidates rendering into a single framebuffer and render pass rather than $K$ separate passes, reducing render target switching overhead. The scene management and object manipulation remain CPU-bound, further highlighting the M2's unified-memory advantage over discrete GPUs.

Adding CUDA-OpenGL interoperability eliminates the PCIe bottleneck for frame readback on NVIDIA hardware. This yields throughput increases of 1.35-1.65x on the RTX 4090 and A100, bringing them closer to parity with the M2. More critically, removing the CPU-GPU transfer bottleneck enables the subsequent optimization to fully utilize GPU capabilities, without being constrained by host-side bandwidth.

The most substantial gain comes from hardware instancing with shader-based batched control. By moving instance manipulation from Python loops to GPU-parallel tensor operations, the M2 improves by 5×, while the RTX 4090 and A100 achieve dramatic 107× and 150× speedups, respectively. This demonstrates the successful shift from a CPU-bound to a GPU-bound rendering workload.

Finally, multi-processing adds marginal improvements for the A100 and M2, which are already GPU-saturated. However, it provides an additional speedup of 1.38 for the RTX 4090, suggesting that its CPU-side driver overhead was still a minor bottleneck that could be parallelized.

We next compare PyBatchRender's performance against other popular libraries that provide high-throughput rendering for Python-based RL: Isaac Lab, Maniskill, and Madrona MJX. We use the CartPoleBalance environment with pixel observations across a range of resolutions.

All benchmarks are run on an Nvidia RTX 4090. For a clear, GPU-focused comparison, PyBatchRender is run in a single process (i.e., without the multi-processing optimization). The performance data for competitor libraries is taken from \cite{zakka_mujoco_2025}.

\begin{figure}
    \centering
    \includegraphics[width=1\linewidth]{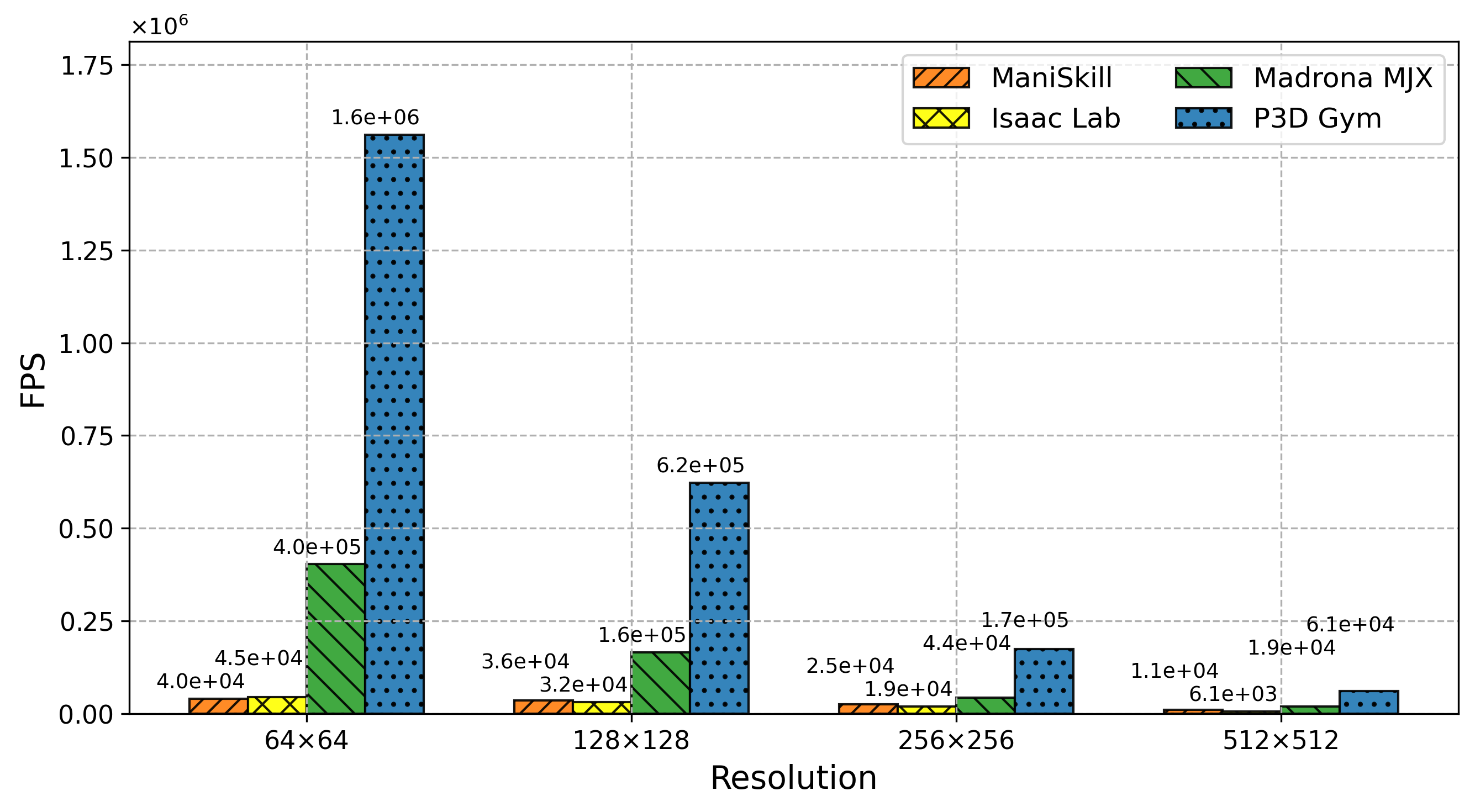}
    \caption{Comparison of environment rendering throughput for CartpoleBalance with pixel observations on an Nvidia RTX 4090. PyBatchRender (without multiprocessing) is compared against Isaac Lab, Maniskill, and Madrona MJX across various resolutions. Competitor data is sourced from \cite{zakka_mujoco_2025}.}
    \label{fig:comp}
\end{figure}

The results, shown in Figure~\ref{fig:comp}, demonstrate that PyBatchRender maintains a substantial performance lead over all compared libraries across all tested resolutions. Isaac Lab and Maniskill share similar performance profiles, with Maniskill performing better at higher resolutions. PyBatchRender significantly outperforms its closest competitor, Madrona MJX. At 64×64 PyBatchRender achieves 1.6 million FPS compared to Madrona MJX's 403k FPS. This gap persists across resolutions, narrowing slightly at 512×512 where PyBatchRender renders at 61k FPS versus Madrona MJX's 19k FPS (3.2× faster).

\section{Discussion}

PyBatchRender achieves over 1 million FPS while maintaining pure Python accessibility, demonstrating that high-throughput rendering need not require building specialized engines from scratch. By extending an existing mature framework with targeted optimizations, we achieve competitive performance with significantly reduced complexity. However, these results must be contextualized within rendering technology trade-offs and practical training pipeline constraints.

PyBatchRender's rasterization-based architecture represents a deliberate design choice with important trade-offs. As demonstrated by \cite{rosenzweig_high-throughput_2024}, ray tracing can outperform rasterization when three conditions converge: low resolution, high polygon counts, and data-center GPUs with a low number of cores specialized in rasterization. However, this performance crossover region has limited practical relevance. Low-resolution observations cannot capture the geometric detail that high-polygon meshes provide. An alternative approach involves level-of-detail techniques that reduce mesh complexity proportionally to resolution, where rasterization remains highly competitive. Additionally, ray tracing's advantages diminish as resolution increases toward the 128×128–256×256 regime increasingly common in modern visual RL \cite{zhou_autoeval_2025, kim_fine-tuning_2025}.

While PyBatchRender achieves over 1 million FPS, rendering performance must be contextualized within the complete training pipeline. On an RTX 4090, optimized Proximal policy optimization (PPO) implementations process approximately 30k-50k frames per second \cite{zakka_mujoco_2025}. Once rendering exceeds this threshold, it ceases to be the primary bottleneck; further 10× improvements in rendering speed yield diminishing impact on end-to-end training time. This observation informed our design philosophy: beyond adequate performance (50k-100k FPS), effort is better invested in interface simplicity, feature richness, and ecosystem integration rather than pursuing extreme throughput.

Several extensions for the PyBatchRender environment are to be implemented in the near future. They include native animated skeletal meshes, dynamic shadow mapping, and advanced per-instance culling. While Panda3D provides these capabilities, naive integration would degrade performance by reintroducing per-object overhead. The library currently targets PyTorch for its flexibility and ecosystem dominance, enabling seamless TorchRL integration and efficient multi-process parallelism. However, the JAX ecosystem has gained significant traction in RL through Brax, MuJoCo MJX, and JAX-based training implementations. Future work should consider JAX support to enable tighter integration with these physics simulators and potentially leverage JIT compilation for lower-overhead environment stepping.

The central motivation behind PyBatchRender extends beyond raw performance metrics. By combining high throughput with Python accessibility, we aim to democratize pixel-based RL. Historically, competitive performance in pixel-based RL required either systems programming expertise or acceptance of communication bottlenecks from engine-wrapper architectures. PyBatchRender provides performance comparable to specialized C++ systems while maintaining rapid prototyping cycles, as researchers can specify custom environments in 30-100 lines of pure Python. The physics-agnostic design further supports diverse research directions, allowing teams to adopt high-performance rendering without committing to particular physics backends. This modularity and accessibility enable smaller research groups and individual researchers to conduct large-scale visual RL experiments previously practical only for well-resourced labs.

% \textbf{Measure how certain features affect the speed}

% \textbf{Compare with:}
% \begin{itemize}
%     \item Madorna 
    
%     \item MuJoCo on JAX (Madorna rendering)
%     \item Isaac Lab
%     \item Maniskill
    
%     \item Godot RL
%     \item Unity RL
%     \item Unreal RL

%     \item Galactic
%     \item Large Batch Rendering?
%     \item Megaverse

%     \item Habitat
%     \item PufferLib
%     \item Megadrive
%     \item DoomViz
%     \item PixelBrax
% \end{itemize}

% \textbf{Figures and Tables }
% \begin{itemize}
%     \item Benchmark table
%     \item Environments frames
% \end{itemize}

\bibliography{p3d_gym}

\end{document}